\documentclass[manuscript]{emulateapj}
\usepackage{graphicx}
\usepackage[normalem]{ulem}
\usepackage{enumitem}
\usepackage{color}

\shorttitle{ Merger history of central galaxies}
\shortauthors{M. Raouf et al.}

\definecolor{darkgreen}{rgb}{0.0,0.5,0.0}
\definecolor{darkred}{rgb}{0.5,0.0,0.0}
\definecolor{brown}{rgb}{0.65,.16,0.16}
\definecolor{grey}{rgb}{0.4,0.5,0.6}

\begin{document}
\title{ Merger history of central galaxies in Semi-Analytic Models of galaxy formation}


\author{Mojtaba Raouf,\altaffilmark{1} 
	Habib G. Khosroshahi,\altaffilmark{1,2}
	Gary A. Mamon,\altaffilmark{2}
	Darren J. Croton,\altaffilmark{3}
	Abdolhosein Hashemizadeh,\altaffilmark{4}
    Ali A.~Dariush\altaffilmark{5}
	}

\affil{$^1$School of Astronomy, Institute for Research in Fundamental Sciences (IPM), Tehran, 19395-5746, Iran}	
\affil{$^2$Institut d'Astrophysique de Paris (UMR 7095: CNRS \& Sorbonne Universit\'e), 98 bis Bd. Arago, F-75014 Paris, France}	
\affil{$^3$Centre for Astrophysics \& Supercomputing, Swinburne University of Technology, PO Box 218, Hawthorn, Victoria 3122, Australia}
\affil{$^4$International Centre for Radio Astronomy Research (ICRAR), The University of Western Australia, 35 Stirling Highway, Crawley, WA 6009, Australia}
\affil{$^5$Institute of Astronomy, University of Cambridge, Madingley Road, Cambridge CB3 0HA, UK}

\email{*m.raouf@ipm.ir}

\begin{abstract}
{ We investigate the dynamical evolution of galaxies in groups with
  different formation epochs. Galaxy groups have been selected to be in
  different dynamical states, namely dynamically old and dynamically young, which
  reflect their early and late formation times, respectively, based on their
  halo mass assembly. Brightest galaxies in dynamically young groups have
  suffered their last major galaxy merger typically $\sim 2$ Gyr more
  recently than their counterparts in dynamically old groups.  Furthermore,
  we study the evolution of velocity dispersion in these two classes and compare them with the analytic models of isolated
  halos. The velocity dispersion of dwarf galaxies in high mass, dynamically young groups increases slowly in time, 
  	while the analogous dispersion in dynamically old high-mass groups is constant. In contrast, the velocity dispersion of
  	giant galaxies in low mass groups decreases rapidly at late times. This increasing velocity bias is caused by dynamical friction, and
  	starts much earlier in the dynamically old groups. The recent {\sc Radio-SAGE} model of galaxy formation suggests
  that radio luminosities of central galaxies, considered to be tracers of
  AGN activity, are enhanced in halos that assembled more recently, independent of the time since the last major merger. }

\end{abstract}

\section{Introduction}

In the hierarchical structure formation framework, massive galaxy systems,
such as clusters, are formed through the mergers of smaller mass systems, such
as galaxy groups. This halo growth is expected to have consequences on the
constituent galaxies. In an isolated halo, dynamical
friction \citep{Chandrasekhar1943} should cause  galaxy orbits to decay toward the center of groups
over a time-scale of a few Gyr \citep{Jones2000,Tollet+17}. As a result, given
sufficient time, the galaxies within the core of the halo, e.g.
$r < 0.5\,r_{\rm vir}$, would merge and form a significantly more luminous
\citep{Ostrik75} and massive galaxy, leaving a substantial luminosity gap between the brightest group galaxy (BGG) and the second brightest. 

Such systems, dubbed as fossil groups, have been identified in
simulations, where it was shown that the luminosity gap is a simple indicator
of the
the frequency of galaxy mergers 
\citep{Schneider1982,Mamon1987}.
Moreover, 
  by analyzing the evolution of the halo mass of fossil groups in
  cosmological dark matter simulations, \cite{Don05} showed that more than
  50\% of the present-epoch halo mass of fossil groups assembled
  at $z\gtrsim 1$, through growth  by minor mergers. Following this pioneering
study, \citet{Dar07} showed
that a luminosity gap of $\sim$2 magnitudes at the current epoch would mean a
$\sim$25\% more massive progenitor at $z$=1, than average. Other studies have  looked for more efficient ways of identifying early formed halos. 
 \citet{Sales2007} and \citet{Dariush2010} investigated the luminosity gap between
the first, 4th and 10th luminous galaxy in the group. \citet{Raouf2014} took this a step
further and showed that a combination of the luminosity gap and a measure of the halo
relaxation, probed by the off-set between the BGG and the group luminosity
centroid, is even more efficient in distinguishing the early and late formed
halos.

The rate of galaxy mergers in groups and clusters is known to scale as
  $1/\sigma_v^3$, where $\sigma_v$ is the group/cluster velocity dispersion,
  both in direct mergers
  \citep{Mamon92,Makino_Hut97,Krivitsky_Kontorovich97}, and as a result of
  dynamical friction (e.g., \citealp{Jiang+08}).\footnote{\label{fn:dfricrate}The dynamical friction merger rate between subhaloes was calibrated with hydrodynamical simulations by \cite{Jiang+08}, to yield a rate proportional to $\ln(1+M_1/M_2)/(M_1/M_2)$, which is roughly proportional to $1/M_1$, hence to $\sigma_v^{-3}$.}
This suggests that the low velocity dispersion is the
  driving parameter for building large magnitude gaps. Since the circular velocity of haloes in $\Lambda$CDM simulations  is roughly constant
 from 7 Gyr ago ($ z \sim 1 $) (e.g. \citealp{Mamon+12, Tweed+18}), one expects that the observable redshift evolution of group velocity
 dispersions can be used to test effects of mergers in groups of different dynamical ages.
 In the present article, we track the redshift evolution of velocity dispersion for  galaxy groups of different dynamical ages and
 compare with the expected analogous evolution of the dark matter particles in their parent halos.

Other studies have added detail to this picture. \citet{Von08} presented evidence for the connection between a large
luminosity gap in fossil groups and the early infall of massive
satellites due to their earlier halo formation with respect to the normal
groups. They found that the time since the last major merger for a majority of
fossils was over 3 Gyr. Recently, \citet{Kundert2017} showed that large
luminosity gap ($\Delta M_{12} \geqslant 2$) $z = 0$ systems assembled most of
their mass before $z \sim 0.4$ (4 Gyr ago) and have a lack of recent mass accretion, in
comparison to groups with $\Delta M_{12} < 2$.

More generally, the properties of a BGG may be related to its luminosity gap and thus
highlight differences in the group assembly history.
For exmple, BGGs that sit in the cores of  large luminosity 
gap groups/clusters tend to have non-boxy isophotes \citep{Khosroshahi2006a,Smith2010}, compared 
to those of similar mass in small luminosity gap groups/clusters.
However, there are no significant differences in 
the
stellar populations of  BGGs in observed groups with large and small
luminosity gaps \citep{LaBarbera2009, Trevisan2017}.
Moreover, analyzing
hydro-simulations, \citet{Cui11} found no trend of luminosity gap with age,
metallicity, colour, concentration or mass-to-light ratio.
They also found that the number density of satellite
galaxies in fossils are similar to those in normal groups. They concluded that fossils are
transient phases in the evolution of ordinary galaxy groups.

Recently, \citet{Khosroshahi2017} discovered a significant difference in the
AGN activity of the BGGs probed by their radio emission at 1.4 GHz.
They found that the BGGs in dynamically relaxed groups (old) are less radio
luminous compared to the BGGs of similar mass residing in dynamically
unrelaxed groups (young). This suggests that AGN activity in BGGs is
influenced by the  dynamics of the group.
This follows earlier clues from studies of a limited sample of confirmed fossil
galaxy groups, which indicated a relatively low radio emission associated
with the fossil BGGs \citep{Jetha08, Miragh14,Miragh15}.
Accretion onto the
super-massive black hole in a galaxy that has experienced its last major
merger a few Gyr earlier is expected to be lower due to the lack of significant
new mergers or tidal effects, itself caused by the absence of massive neighboring
galaxies in large luminosity gap systems. However this does not rule out other forms
of accretion, such as hot mode accretion. That said, the epoch of the last major merger
may determine whether radio emission is boosted in dynamically young galaxy groups. By exploiting the Illustris
hydrodynamical simulation, we showed that the central supermassive
black hole in an old group is more massive than one in a young
group, and that black hole accretion occurs at a slower rate \citep{Raouf2016}.
\begin{figure}
	\centering
	\includegraphics[width=0.49\textwidth]{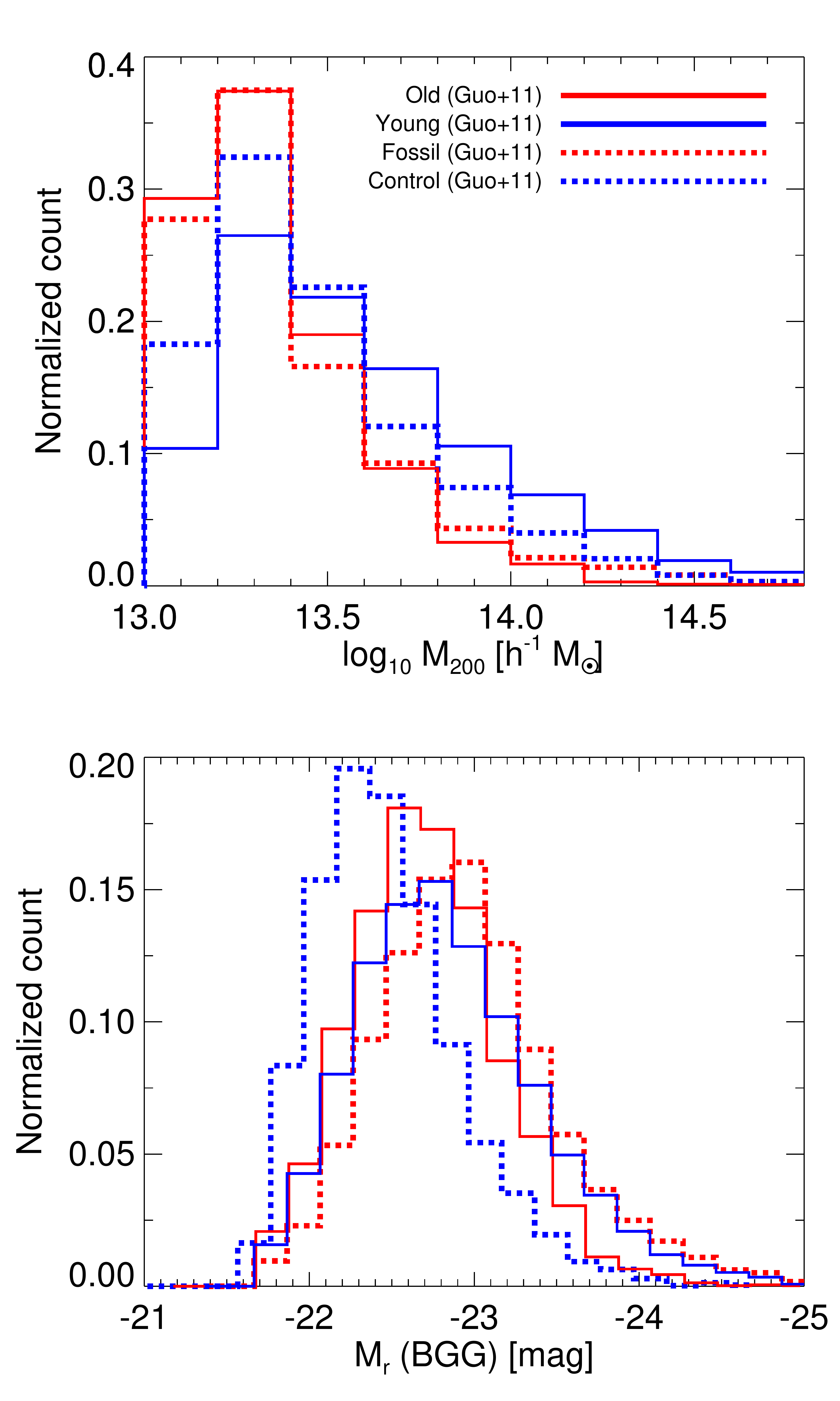} 
	\caption{Distributions of $z$=0 halo mass (\emph{top}) and BGG r-band absolute magnitude (\emph{bottom}) for dynamically old (\emph{red line}),
		young (\emph{blue line}), fossil (\emph{red dashed line})
		and control (\emph{blue dashed line}) galaxy groups
		identified in the \cite{Guo11} semi-analytic model.
	}
	\label{fig:Mass_Guo}
\end{figure}

Fossil groups have also been scrutinized by other means such by their halo
concentration \citep{Khosroshahi2004a,Khosroshahi2007,Buote2016} and X-ray
properties \citep{Jones2003, Khosroshahi2006b}. Significant steps have been taken to understand the nature of fossil groups
\citep{Aguerri2011}, although sample selection and a possible contamination
in the samples have resulted in some contradictions.  For instance, \citet{Kundert2017} found no difference in the
group mass formation at redshift $\sim 1$ due to a conservative sample
selection for comparison of fossil ($\Delta M_{12} \geqslant 2$) and
non-fossil ($ \Delta M_{12} < 2$) groups.
\begin{figure*}
	\includegraphics[width=1.1\textwidth]{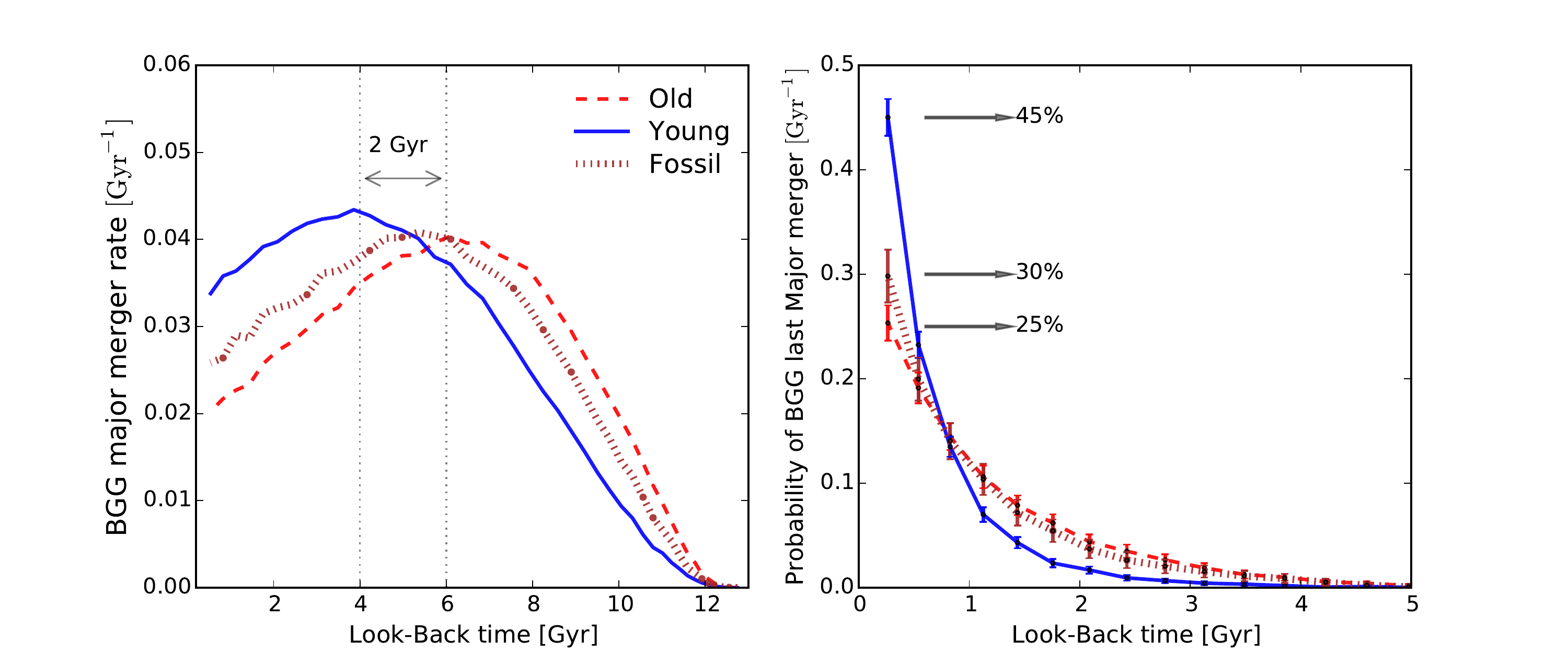}
	\caption{{\bf Left:} Merger history of the BGG in three categories of
		groups: dynamically old (\emph{red long-dashed line}) and young
		(\emph{blue line}) groups of galaxies along the fossil
		($\Delta M_{12}>2$) groups (\emph{brown dotted line}), for halo masses above
		$10^{13} ~ h^{-1}$ M$_\odot$. Major mergers are characterized by mergers
		with $m_1/m_2 \leq 3$. As seen, the BGGs of old systems shows a
		higher probability for a major merger at earlier epochs compared with
		young and fossil systems (about 2 Gyr in peak). 
		{\bf Right:} Comparison of the distributions of time since the last
		major merger between the three group
		categories. The error bars are Poisson. The last BGG merger in young
		groups is a more recent event than the same
		for the BGGs in fossil and old groups, respectively.
}
	\label{BGG-merger}
\end{figure*}

Cosmological simulations have helped us understand the halo-galaxy connection, while observations of galaxy properties
in groups, and in particular where a distinction is made according to dynamical state, have allowed us to probe the success of
galaxy formation models, implemented semi-analytically or hydro-dynamically
\citep{Raouf2014, Raouf2016, Gozaliasl2014, Gozaliasl2016}. In this tradition, here we revisit the merger history of the dominant galaxy. We study
the impact of the last major merger of central galaxies within dynamically old
and young groups of galaxies on their associated 1.4 GHz AGN radio emission, as predicted
by the {\sc Radio-SAGE} semi-analytic model of galaxy formation. We take a closer look
at galaxy velocity dispersions in groups and its evolution, and compare it with analytical predictions. 

Today's large-scale cosmological dark matter
simulations, coupled with Semi-Analytical Models (SAMs) of galaxy formation,
have made it possible to track the properties of groups and clusters of
galaxies backward in time and study their evolutionary histories. In this
way, we can measure and trace back the physical properties of galaxies, which is otherwise impossible to do
observationally. We utilize the Millennium Simulation \citep{Springel2005},
one of the largest cosmological dark matter simulations ever performed, together
with the semi-analytical galaxy catalogues of \citet{Guo11} and our {\sc Radio-SAGE} model \citep{Raouf2017}. This allows us to
identify and study fossil and control groups of galaxies, as well as old and young
galaxy systems representing the relaxed and unrelaxed galaxy groups, as
introduced in \citet{Raouf2014}.

The structure of the paper is as follows: the data and sample selections are
described in Section \ref{sec:Mill}.  In Section \ref{sec:results1}, we
present our results and analysis. Finally, we provide a summary and
discussion in Section \ref{subsec:sum}. Throughout this paper, we adopt 
$H_0 = 100 h\,\rm km\,s^{-1}\, Mpc^{-1}$ for the Hubble constant with
$h=0.73$.

\section{Data and sample selection} \label{sec:Mill}

\subsection{Millennium Simulation}

In this study, we used the Millennium Simulation (hereafter, MS,
\citealp{Springel2005}),
containing $2160^3$ particles with a mass resolution of $8.6 \times 10^{8}\, h^{-1} {\rm M_{\odot}}$
within the box size of $(500\, h^{-1}\rm{Mpc})^3$. The simulation begins at
redshift $z$=127 and evolves up to the present epoch and stores the data in 64
separate snapshots. The epochs of interest to us are $z<1$, taking
into account the importance of the last $\sim$7 Gyr in the evolution of
galaxy groups, as well as the limitations of the simulation in terms of the
halo mass and a number of group members. The MS assumed cosmology parameters as described in {\it Wilkinson Microwave
  Anisotropy Probe -1} \citep[WMAP-1;][]{Sper03}. Dark matter halos were
found by means of a friends-of-friends (FoF) algorithm as described in
\citet{Dav85}. The SUBFIND algorithm \citep{Spr01} was then applied to the
FoF catalog in order to identify subhalos by restricting boundaries of
substructures. The halo mass resolution in the MS is 20 particles,
i.e. a minimum halo mass of $1.72 \times 10^{10} \,h^{-1} {\rm M_{\odot}}$.

\subsection{Semi-Analytical Models}

Halo merger trees provided in the publicly available German Astrophysical
Virtual Observatory (GAVO)
database\footnote{http://gavo.mpa-garching.mpg.de/MyMillennium}
allow us to link halos and
sub-halos
between time-steps
(snapshots).
Semi-Analytical Models (SAMs) follow the many branches of the merger tree from past to
present with specific recipes for galaxy properties, such as the gas, and
stars of the bulge and disk components of galaxies
\citep{Croton2006,Bow06,DeL07, Guo11,Raouf2017}.
In this study,  we use the publicly available semi-analytical catalogue of
\cite{Guo11}, which is successful in predicting the observed luminosity and stellar mass functions of galaxies, from the SDSS 
data and recent determinations of the abundance of satellite galaxies around the Milky Way as well as the clustering properties of galaxies as a function of stellar mass by updating the number of physical processes such as galaxy morphology and environmental effects and the treatments of the transition between the central and satellite galaxies during merger.
 Our sample includes  $\sim 39\,000$ halos of mass
$M_{200}>10^{13} \,h^{-1}\rm M_{\odot}$ with at least four member galaxies
brighter than $M_r = -14$ at $z= 0$. In Sections~3.1, 3.2, and 3.3, we use this
  semi-analytical model 
  to study the merger history, the  evolution of the group velocity
  dispersion (traced by its galaxies) and of the radial distribution of group members.

For comparing the activity of central supermassive black holes in Section 3.4, we use a new galaxy formation model which is a modified
version of the SAGE semi-analytic model \citep{Croton2016} run on the
Millennium simulation \citep{Springel2005}, which we hereafter refer to as {\sc
  Radio-SAGE}
\citep{Raouf2017}.\footnote{https://github.com/mojtabaraouf/sage}
      {\sc Radio-SAGE} incorporates a new method for tracing the
physical properties of radio jets in massive galaxies, including 
the evolution of radio lobes and their impact on the surrounding gas. In our
model, we self-consistently trace the cooling-heating cycle that
significantly shapes the life and death of many types of galaxies.
We compute the radio luminosity, which is an important observable quantity to
study AGN, through the radio luminosity function, the calibration of the jet
power and the environmental effects on AGN.
For self-consistency, we first updated the hot
halo gas density profile into which the jets propagate. This affects the hot
gas cooling rates into the galaxy. We make predictions for intermittent
triggering and radio emission from AGN jets and construct the radio
luminosity function at the same time as the usual stellar mass
functions. The use of two semi-analytic models in this study is
  driven by data products available from these simulations. The details of
  the \cite{Guo11} SAM are not in conflict with those of {\sc Radio-SAGE}
  within the scope studied here.

\subsection{Sample Selections}  \label{sec:definitions}

A useful parameter used in this study, as a proxy for halo mass assembly, is
the ratio of the halo mass at any given redshift $z$ to its final mass at
$z=0$, i.e. $\alpha_{z,0} \equiv M_{z}/M_{z=0}$. In our earlier study it was
argued that a sample of fossil groups identified based on the luminosity gap,
$\Delta M_{12}$, will result in a contaminated sample if the objective was to
form a sample of early formed halos. However, by adding a constraint on the
separation between the group luminosity weight and the position of its BGG,
$D_{\rm off}$, we considerably improved the age-dating method for classifying relaxed and unrelaxed galaxy groups and clusters, in comparison to the method
based on the luminosity gap only \citep{Raouf2014}. There are
other studies suggesting alternative definitions for early formed (relaxed) systems based on the magnitude gap. Accordingly,
\citet{Dariush2010} found 50\% more early formed systems adopting a
magnitude gap definition of $\Delta M_{14} \geq 2.5$~mag as opposed to the
conventional criterion of $\Delta M_{12} \geq 2.0$~mag.

In this work, samples of fossil/control and old/young groups of galaxies are
selected from both the SAM group catalogs of \cite{Guo11} and the {\sc
  Radio-SAGE} (dynamically old and young) model according to the following
criteria:

\begin{enumerate}[label=(\roman*)]
	
	\item {\it Fossil/Control systems}: Fossil groups are defined with
          the conventional definition of a large magnitude gap, $\Delta M_{12} \geq 2.0$~mag \citep{Jones2003}.  We use the
          $r$-band absolute magnitude to 
          estimate the gap within $0.5 R_{\rm 200}$ of halos, where
          $R_{\rm 200}$ is the radius in which the mean  mass
          density is 200 times the critical density, $\rho_{\rm c}(z)$.
          Observationally, fossil groups have a minimum bolometric X-ray
          luminosity of $L_{\textnormal{x,bol}} \geq 0.25 \times 10^{42}\,
          h^{-2}\rm erg \, \rm s^{-1}$, which according to the Millennium gas
          simulation, almost corresponds to groups with halo masses of
          $M(R_{\rm 200}) \gtrsim 10^{13}\, h^{-1} \rm M_{\odot}$
          \citep{Dar07}. Furthermore, fossil groups possess a giant
          elliptical galaxy at the center of their halos. To take this into
          account, our sample of fossil groups are limited to those having
          BGG absolute $r$-band magnitudes of $M_{r} (\rm BGG) \leq
          -21.5$~mag. In addition, when comparing fossils with normal groups,
          we define control groups of galaxies as systems with $\Delta M_{12}
          < 0.5$~mag but with the same halo mass and BGG luminosity as in
          fossil systems. Based on this criteria, we find 6938
          ($\sim 18\%$) fossil and 6053 (15\%) control groups in the
          \citet{Guo11} SAM.
          
	\item {\it Old/young systems}: According to \cite{Raouf2014}, old
          galaxy systems have assembled more than 50\% of their final halo mass by
          $z\approx1.0$, young galaxy systems have only acquired less than 30\% of their final halo masses by $z=1$. Considering these constraints, together with the same
          limits on BGG magnitudes and halo masses as applied to select
          fossil/control systems, we find 12678 ($\sim 32 \%$) old and 7478
          ($\sim 19\%$) young groups in the \citet{Guo11} SAM. We identified the same
          fraction of old and young systems in the {\sc Radio-SAGE} group
          catalogue.   We find that $\sim$54\% of fossil groups overlap with old systems and only
          $\sim$37\% of control groups are also young systems.
\end{enumerate}
 Figure \ref{fig:Mass_Guo} shows the distributions of halo mass and
   BGG $r$-band luminosities of our selected samples at $z$=0 in the
   \cite{Guo11} SAM.

\section{Analysis and Results} \label{sec:results1}

\subsection{BGG merger history} \label{subset:last_merge}

The formation of giant elliptical galaxies through major mergers of disk
galaxies have been shown in both simulations and observations   \citep{Toomre72,
	Naab09}.  Based on this merger scenario, large galaxies lose angular
momentum due to dynamical friction and fall into the center of groups where
they coalesce to form giant elliptical galaxies
\citep{White1976,Schneider1982,Mamon1987, Ponman1994} that are frequently
observed at the center of fossil groups and clusters of galaxies. In a
scenario presented by \cite{Bar89}, the BGGs of
fossil groups, known to be among the most massive galaxies in the Universe,
are formed through the merging of central bright galaxies in compact galaxy
groups. According to this scenario, compact groups are suggested to be the
progenitors of fossil groups and central giant elliptical galaxies are the final
remnants of group scale mergers.  Using N-body simulations, \citet{Farhang2017} traced the halo
mass of compact groups from $z = 0$ to $z = 1$ and showed that, on average, 55\% of
the halo mass in compact groups is assembled since $z \sim 1$, compared to
40\% of the halo mass in fossil groups over the same time interval. This indicates
that, compared to fossil groups, compact groups are relatively younger galaxy
systems. They found that while $\sim$ 25\% of fossil groups go through a
compact phase, most fail to meet the compact group isolation criterion,
leaving only $\sim$30\% of fossil groups fully satisfying the compact group
selection criteria.

Some studies have found no sign of a recent major merger in
the most luminous galaxy within some fossil groups.  For instance,
\cite{Khosroshahi2006a} and \cite{Smith2010} have pointed out the absence of
ongoing group-group mergers and of major galaxy-galaxy mergers in fossil systems,
based on X-ray studies of the morphology of a sample of regular/relaxed fossil
groups.  Such a result directly points to an early formation epoch in fossil
groups.  

The age-dating method used in this study is based on the mass
assembly history of group halos, as introduced by \cite{Raouf2014}. It is
interesting to see how the BGG merger history is influenced by the mass
assembly of its host halo. To this end, and by using the merger trees of the MS
and tracing the histories of the BGGs, we study the merger history of the BGGs in three types
of groups i.e.,
 the old, young and fossil galaxy groups as described in Section \ref{sec:definitions}.
\begin{figure*}
	\includegraphics[width=1.\textwidth]{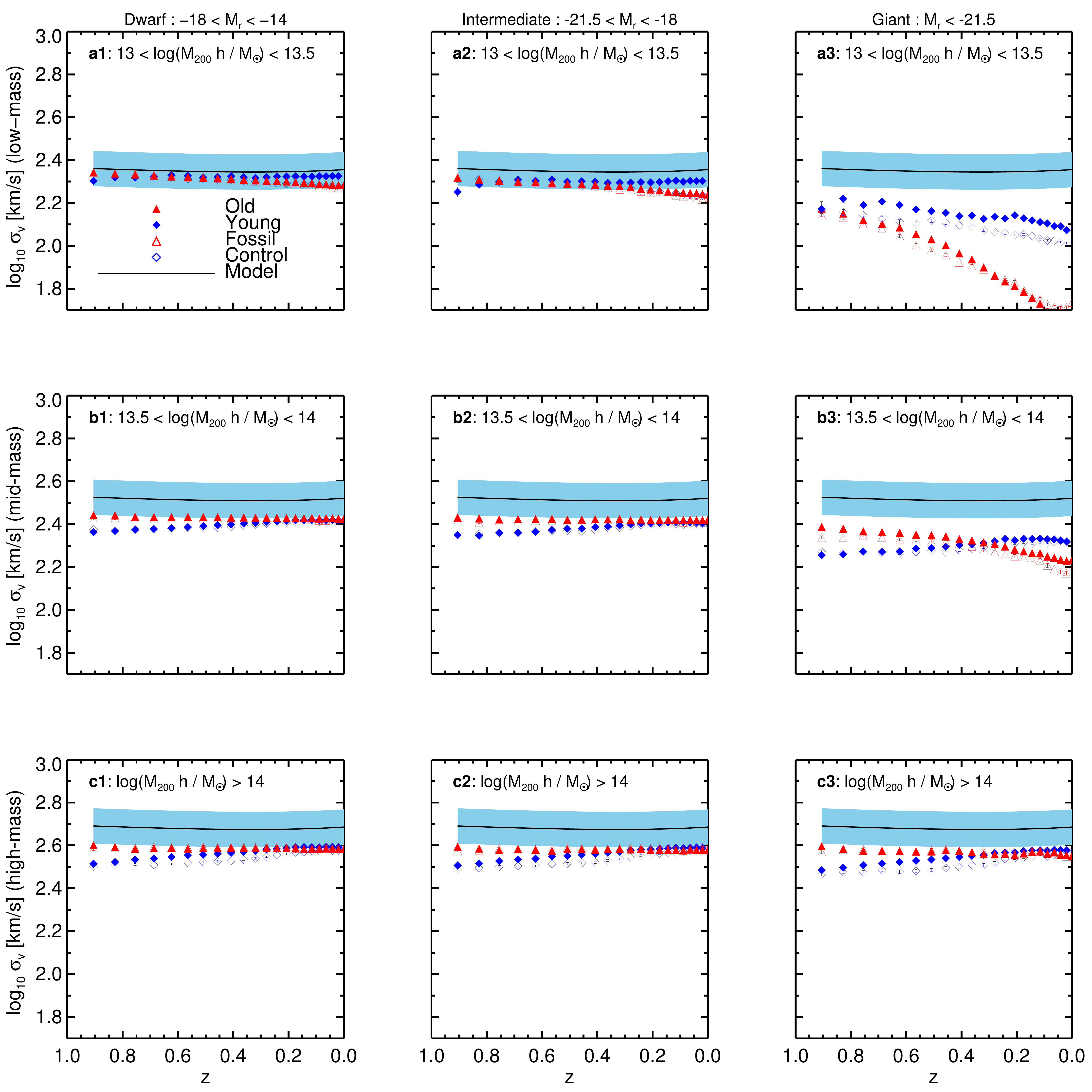}
	\caption{Redshift evolution of the 1D group velocity
			dispersion measured within $R_{200}$ using the dwarf, intermediate and giant galaxies
          (\emph{left}, \emph{middle} and \emph{right columns}, respectively)
          for old (\emph{red filled triangles}) and
          young (\emph{blue  filled diamonds}) groups, as well as 
          fossil (\emph{red triangles}) and control (\emph{blue diamonds})
          groups.
          The symbols show the mean values, and the error bars
          are the uncertainties on the mean ($\sigma/\sqrt{N}$). The
          variation of mean velocity dispersion for the model of isolated
          halos with assuming $f_c \equiv \sigma_v / V_v = 0.65 $
          (see Appendix~\ref{app}),
          are displayed as
          \emph{black curves} and the \emph{blue shaded regions} indicate the
          upper and lower limits with typical 0.25 dex uncertainties for the $z$=0 halo
          mass range. Groups are split into three mass bins from top to
          bottom.
        }
	
	\label{fig2:sigma-v}
\end{figure*}

We only consider groups with halo masses $ \geq 10^{13} \,h^{-1} {\rm
  M_\odot}$. We classify mergers into major and minor based on the mass ratio
$m_1/m_2$ of galaxies involved in merging. A merger is categorized as major
or minor if $m_1/m_2 \leq 3$ or $m_1/m_2 > 3$, respectively. Figure
\ref{BGG-merger} shows probability distributions of this major merger rate
(left-panel) and the time of the last major merger (right-panel) of the central
dominant galaxy, estimated at different epochs in old, young and fossil
groups.  The recent study of \citet{Kundert2017} showed that large gap
($\Delta M_{12} \geqslant 2$) groups assembled less than 20\% of their
mass recently (between $z = 0$ and 0.4), in comparison to normal gap
($\Delta M_{12}< 2$) groups. As can be seen in the left panel, the major merger
phenomenon in BGGs in old (already like fossil) groups peaks at an earlier
epoch ($\approx 2.0$~Gyr) compared to young groups. This indicates that the
evolutionary history of the halos is somehow dictated by mergers. The
right-panel of Figure \ref{BGG-merger} suggests that 45\% of young
systems experience their last major merger before the look-back time of
$\sim$ 0.3 Gyr, while the same in old systems is 25\%. In addition,
more than 45\% of old systems have met their last major merger at
look-back times $\geq 1.0$~Gyr, in comparison with only about 15\% for
young systems. Note that, the predictions for galaxy merger rates can
  differ significantly among theoretical methods \citep[see
    e.g.][]{Hopkins2010,Rodriguez-Gomez2015,Guo2008}.

\subsection{Redshift evolution of velocity dispersion} \label{subsec:sigma}

Not only do galaxy mergers lead to larger magnitude gaps, but they may affect
the velocity dispersions of groups. Dynamical friction causes the
  dissipation of orbital energy of galaxies, which depends on their location
  in the viral sphere and the group to galaxy mass ratio (see
  footnote~\ref{fn:dfricrate}). In high-mass groups, the dynamical friction
  time scales should be too long to reduce the velocity dispersion in the
  last 7.6 Gyr (i.e. since $z=1$), even for massive galaxies. In contrast, in
  low mass groups, dynamical friction should lead to short orbital decay
  times for intermediate- and high-mass galaxies.

In Figure \ref{fig2:sigma-v}, we study the evolution of line-of-sight velocities dispersion (1D velocity dispersion) in dwarf, intermediate and giant members of old, young, fossil and control groups of galaxies, for different group mass range.
 
 The figure shows that velocity dispersion in young
and control groups, and for all members of massive systems, increases with
time (panels: b1, b2, b3, c1, c2 and c3), suggesting that these groups are
\textit{unrelaxed} systems. This increase in group velocity dispersion in young and control systems may be caused by the high fraction of major mergers of galaxies within these groups at redshifts $z < 1$ (Figure \ref{BGG-merger}). 

While there no noticeable changes in the trend of young and control subsamples velocity dispersion through their evolution with time in the low mass groups of galaxies (panels: a1, a2), we can see a slight decrease in the velocity dispersion of giant members (panel: a3). In contrast, there are no significant variations in the evolution of velocity dispersion of old and fossil galaxy groups, suggesting that these groups are \textit{relaxed} systems, except perhaps for the giant members of low and intermediate mass systems (panels: a3 and b3) where the velocity
dispersion decreases with time due to dynamical friction.

In general, Figure \ref{fig2:sigma-v} shows a wide range of velocity
dispersion fluctuations through redshift in both unrelaxed and relaxed
systems. These findings also show the hierarchical evolution of galaxy groups
through their period of life where young groups, which indeed are
non-virialized, grow up in size and mass, leading to an increase of galaxy
velocity dispersion in such systems. In contrast, old groups are systems
which have been formed following the virialization of young systems in a
hierarchical way, and that their halo dynamics/masses have not
experienced a noticeable change for a long period of time\footnote{Figure 7
  in \citet{Dar07} shows mass assembly of a typical fossil group and a
  control group from the Millennium gas simulation from redshift $z = 1$ to
  0. While the fossil group has already largely been assembled, the control
  group has considerable substructure even at a later epoch }.
Like fossil groups, the velocity dispersion of old groups shows a flat trend
from $z\approx 1$ to $z=0$, while young and control groups display important
evolution in their velocity dispersion. We note that less than 4\% of the groups in our samples have less than or equal to two giant members, which would result in a velocity dispersion which is not useful in practice. The error bars show the small effects of the sampling noise on the estimated velocity dispersions.

\begin{figure*} 
	\begin{center}
		\includegraphics[width=1.\textwidth]{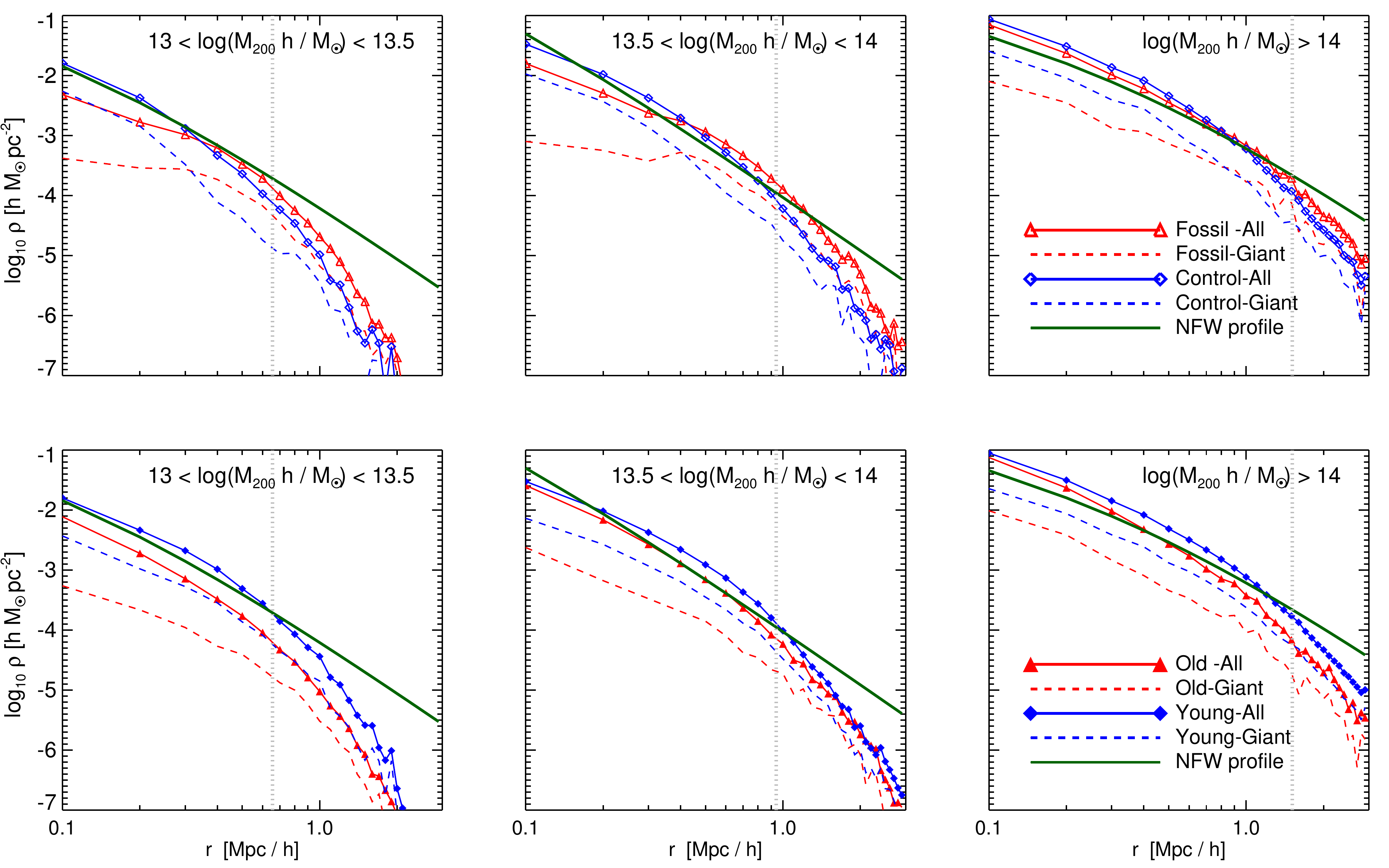}
	\end{center}
	\caption{  Group stellar mass density profiles in fossil
		(\emph{open red triangles}) and control 
		(\emph{open blue triangles}) groups (\emph{top panels}) as well as
		old
		(\emph{red triangles}) and young (\emph{blue triangles}) groups of
		galaxies
		(\emph{bottom panels}). The same distribution for giant
                galaxies in Fossil (\emph{dashed red line}), control
                (\emph{dashed blue line}) groups as well as old (\emph{dashed
                  red line}) and young (\emph{dashed blue line}) groups are
                shown in the \emph{top panels} and  \emph{bottom panels},
                respectively. The poisson error bars in each points are over
                0.5 dex. The NFW profile (Navarro et al. 1996) is  displayed
                (\emph{green line}) for reference. 
		These distributions are shown for 3 bins of $z$=0 halo mass
		increasing from \emph{left} to \emph{right}.
       The sharp declines of stellar mass density beyond the virial
         radii (\emph{grey dotted-lines}) is due to the assignment  of galaxies
         to the FoF halo in the semi-analytical model. 
        }
	\label{fig:R-R200-bright}
\end{figure*}

\begin{figure}
	\begin{center}
		\includegraphics[width=0.45\textwidth]{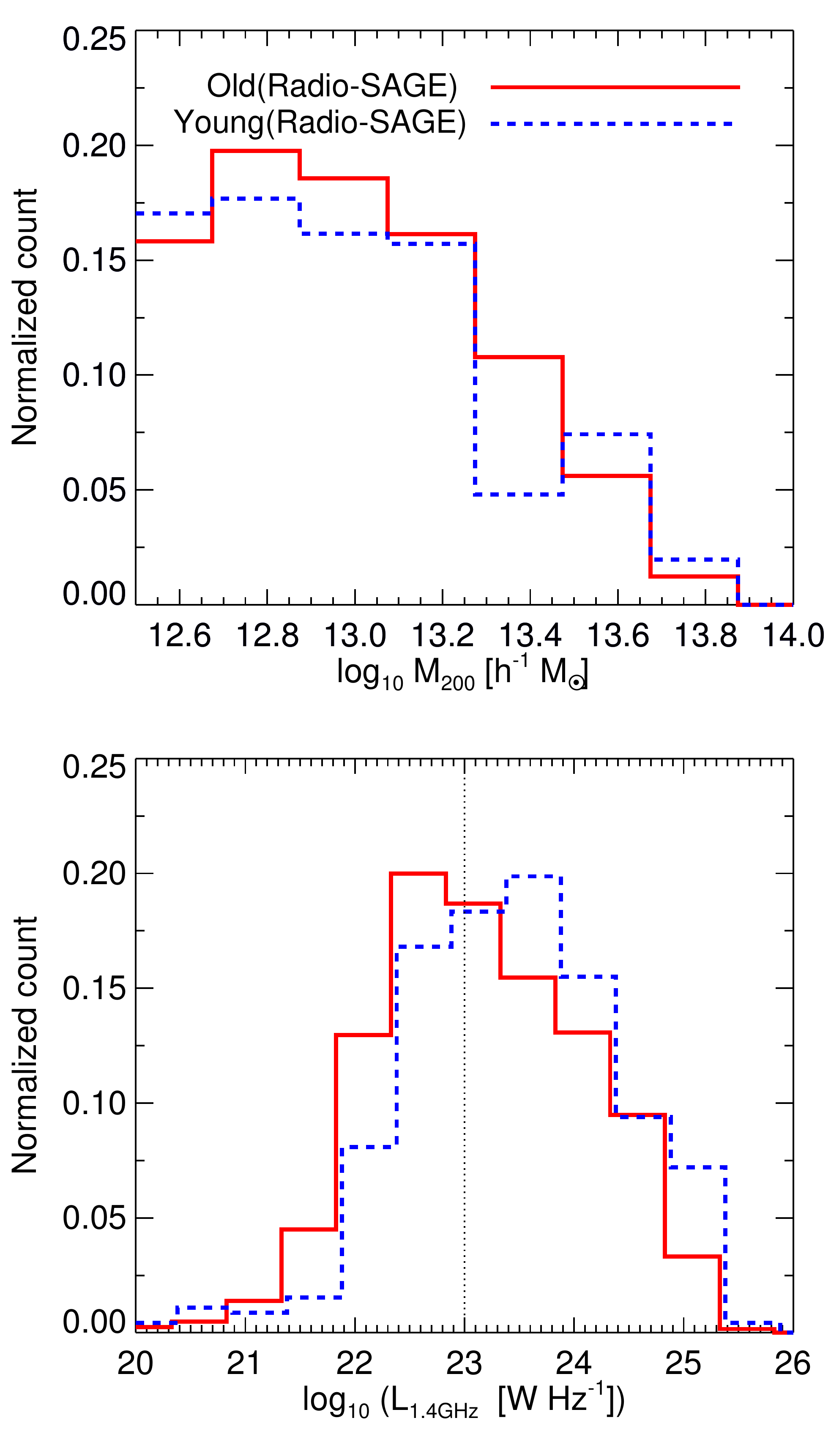} 
	\end{center}
	\caption{Distribution of $z$=0 halo mass (top) and 1.4 GHz radio luminosity (bottom) for dynamically old (\emph{red line})
          and young (\emph{blue dashed line}) galaxy groups identified at $z = 0$ in the {\sc
            Radio-SAGE} semi-analytic model. }
	\label{fig:Mass}
\end{figure}

\subsection{Radial Distribution of Galaxy Group Members} \label{subsec:scaled-radius}

In Figure \ref{fig:R-R200-bright}, we show the radial distribution
  of galaxy group members in fossil/control and old/young groups as a
function of their radial distance to the BGG for various halo mass bins.
The stellar density follows the NFW model \citep{Navarro+96} up to
  the virial radius and becomes steeper at larger radii, because of how the
  galaxies are assigned to their haloes.
The upper panels indicate that the control groups have higher stellar mass density in the inner regions than fossil groups. However, within the same radius and
	compared to young groups, the higher density of galaxies in old systems implies
	are similarly centrally concentrated.

Similar density distributions are shown for giant galaxies defined as those
with $M_r \leq -21.5$. By comparing the mass density of giant galaxies in old
groups and at different radius with those in fossil groups, we find that the
abundance of giant galaxies on the outskirts of old groups is less than
around fossil groups. In addition, a majority of giant galaxies, which mostly
represent the BGG counterparts in old groups, settle at the centers
of halos within $r \leq 0.2\, {\rm Mpc}/h$.  A gap also exists between fossil
and control systems around $0.2 \leq r \leq 0.5\,{\rm Mpc}/h$ (top panels in
Figure  \ref{fig:R-R200-bright}).  This is due to our selection based on the
magnitude gap. Similarly, there is a peak at radii larger than $r \geq 0.5
\,{\rm Mpc}/h$ in the distribution of giant galaxies in fossil groups, again
showing a selection effect in finding relaxed systems based on the
magnitude gap only criterion, due to the lack of $M_*$ galaxies in
fossil groups.  We also note that the BGGs are at the very center in
  all panels of the above figure and do not appear in these plots. Such
selection effects are not observed in genuine old groups as giant galaxies in
such systems have had enough time to fall towards the center of their host
halos through dynamical friction \citep{Von08}. 
The expected NFW model for such a distribution is shown with the black dashed-line in all
panels of the above figure.

\subsection{Radio luminosity  vs. last major merger}
In a recent study, \citet{Khosroshahi2017} used GAMA observations to show that
most massive galaxies in dynamically unrelaxed galaxy groups have
higher AGN accretion rates in comparison to those in dynamically relaxed
galaxy groups. They argue that such an observed phenomenon is due to the
presence of higher jet power in radio wavebands from the brightest
galaxies. These findings also agree with our recent
study of galaxy groups in the Illustris hydrodynamical simulation, in which
we found a lower rate of black hole accretion at a given stellar mass of
old BGGs in comparison to BGGs in young galaxy groups \citep{Raouf2016}.

 Here, we study how radio luminosity in dynamically old and young groups varies with stellar mass and other quantities using
the {\sc Radio-SAGE} semi-analytic model. The top panel of Figure \ref{fig:Mass} shows their
  distribution of halo masses. In the bottom panel, we show that the younger systems ($\sim 70\%$) are in fact more radio-loud (by $\approx$ 0.4 dex) than the older systems($\sim 57\%$)  in this model.

\begin{figure}
	\begin{center}
		\includegraphics[width=0.5\textwidth]{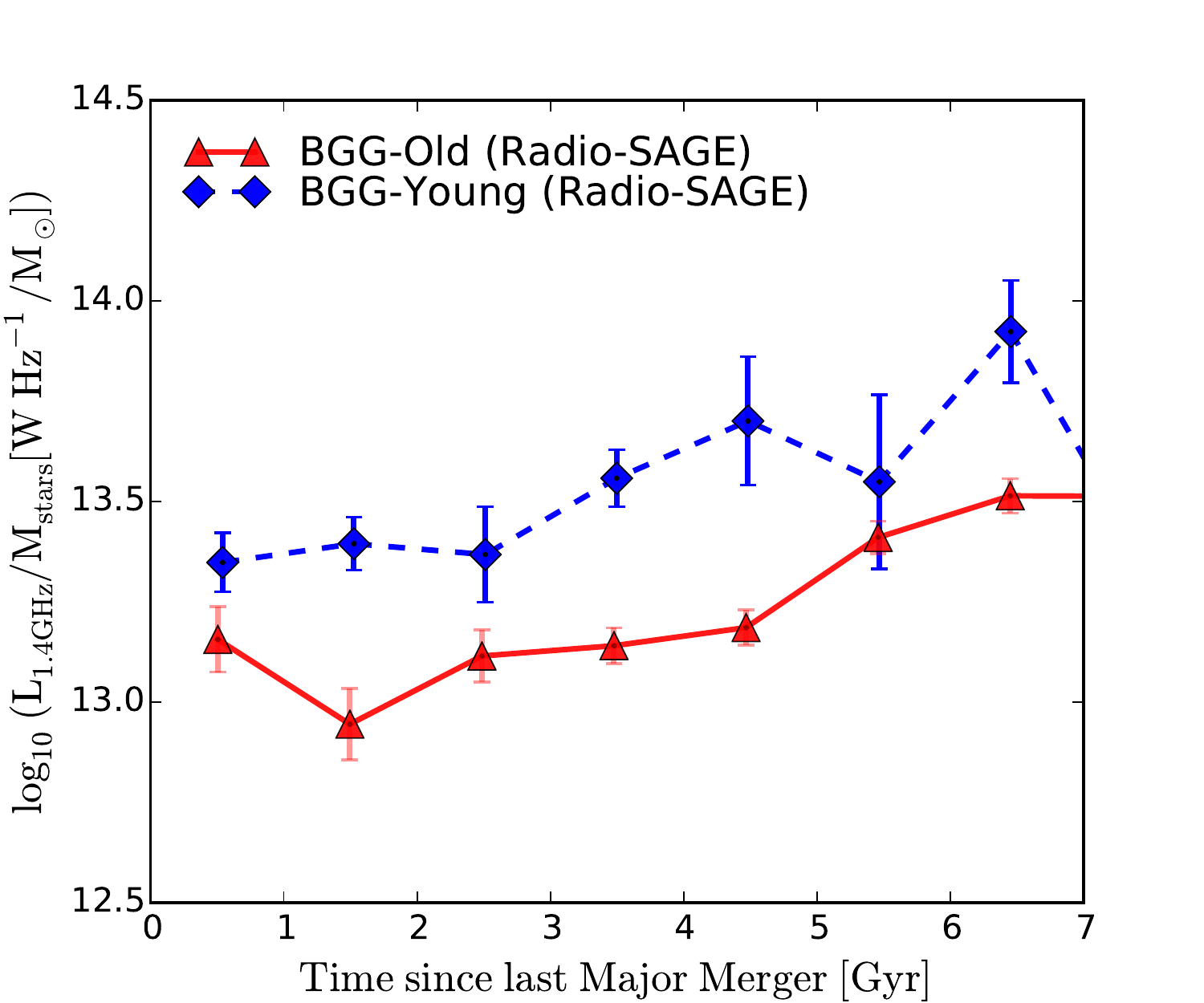}
	\end{center}
	\caption{ Radio loudness (luminosity at 1.4 GHz divided by stellar
		mass) of BGGs at $z=0$ versus the time since the last major merger suffered
		by the BGG,
		in subsamples of groups defined by the epoch of their
		last major merger. The \emph{red triangles} and \emph{blue diamonds} show the
		medians and $\sigma/\sqrt{N}$ uncertainties for BGGs of old and
		young groups, respectively.}
	\label{fig:SAGE_Merger}
\end{figure}

\begin{figure}
	\begin{center}
		\includegraphics[width=0.55\textwidth]{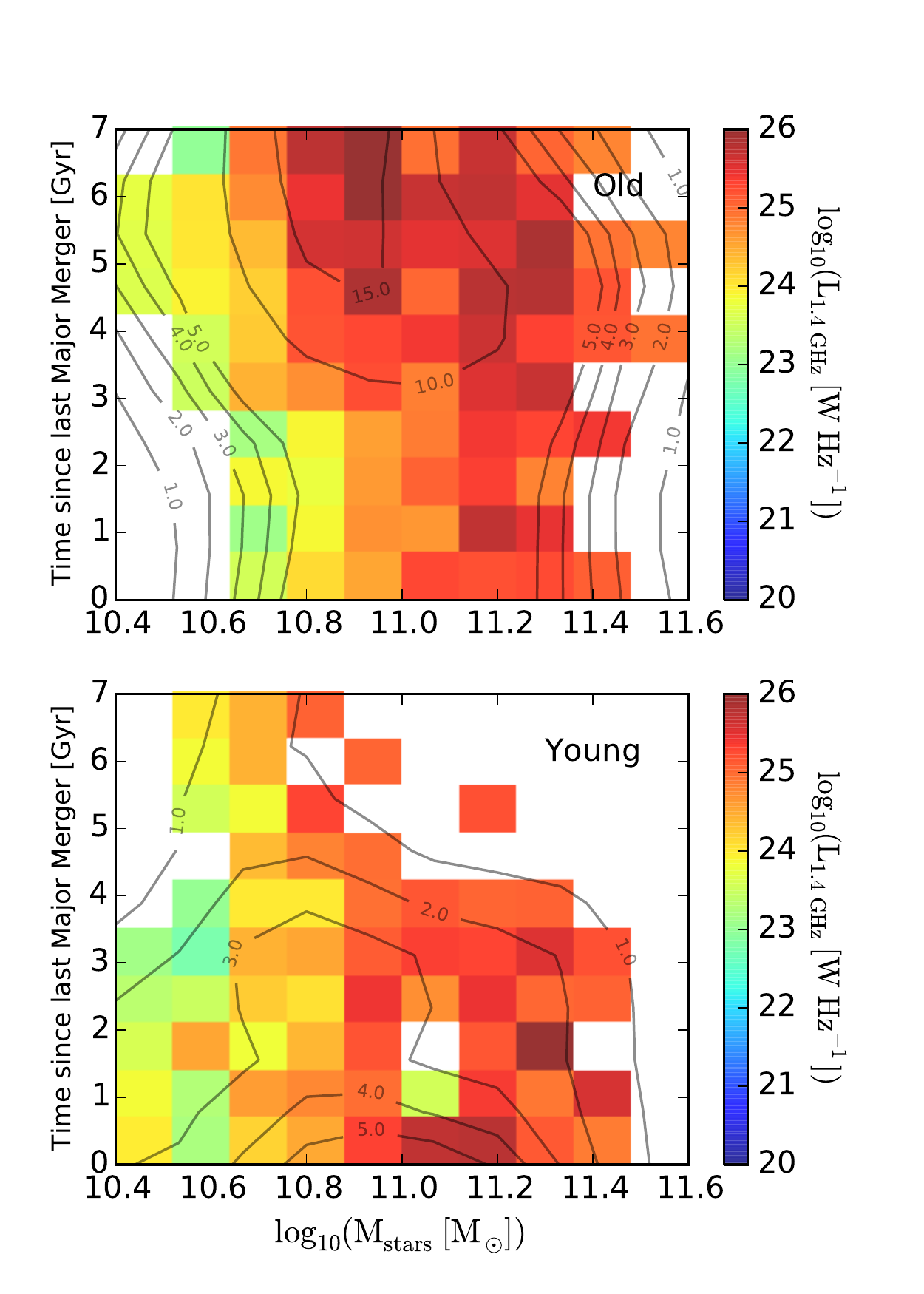}
	\end{center}
	\caption{Time since the last major merger
		suffered by the BGGs versus their
		stellar mass for the BGGs of old (top panel) and
		young (bottom panel) groups, with symbols color-coded by the 1.4 GHz radio
		log luminosity, as derived from the {\sc Radio-SAGE} SAM.
		The \emph{black contours} show the density of galaxies per pixel (after smoothing by a Gaussian with $\sigma$ = 1.0 pixel).
	}
	\label{fig:SAGE_Merger_mass}
\end{figure}

 So radio luminosity is enhanced in groups that have assembled (merged) recently. To understand how this impacts the BGGs, we compare, in Figure \ref{fig:SAGE_Merger}, the radio loudness of BGGs in young and old groups as a function of the time since the last major merger (hereafter TMM, ratio of stellar masses from 1:1 to 0.3:1) suffered by the BGG. The figure indicates that the radio luminosity of BGGs in young groups is enhanced by a factor 2 relative to BGGs in old groups of the same time since the last major merger.
Moreover, one sees that radio loudness increases slightly with increasing TMM. This result seems surprising, given that, in 
{\sc Radio-SAGE},  galaxy mergers lead to gas fueling of the AGN, and then to enhanced radio power. However, as can be seen in 
Figure \ref{fig:SAGE_Merger_mass}, it is not so much that radio luminosity increases with increasing TMM, but rather the stellar mass decreases with TMM (see the anticorrelation of stellar mass and TMM in the black contours of Figure \ref{fig:SAGE_Merger_mass}, for both old and young groups). 

Note that {\sc Radio-SAGE} predicts roughly similar trends of 1.4 GHz radio luminosity versus stellar mass, separated between old and young groups, as are observed \citep{Khosroshahi2017}. 
Our previous analysis of the ILLUSTRIS hydrodynamic simulation has shown the higher AGN accretion
	rate of BGGs hosted by dynamically young groups compared to the old
        systems at a given BGG stellar mass \citep{Raouf2016}, as confirmed
        with the observational study of \cite{Khosroshahi2017}. Here, we show
        that the BGGs in late formed systems have higher 1.4 GHz radio
        emission, independent of the elapsed time since its last major
        merger.

\section{Discussion and Summary}
\label{subsec:sum}

This paper focuses on the merger history of the brightest group galaxies in
dynamically old and young galaxy groups to understand
the connection between their evolution and that of their host halos. Our study
is primarily limited to group size halos. We apply two semi-analytic models
of galaxy formation to follow the BGGs: the Munich model \citep{Guo11} and the newly
developed {\sc Radio-SAGE} model \citep{Raouf2017}, both 
run on the Millennium Simulation.
In particular, {\sc Radio-SAGE} allows a more accurate analysis of the AGN
activity of such galaxies.
In our study, we make a
distinction between fossil/control galaxy groups and dynamically old/young
groups. The fossil/control distinction is driven by the luminosity gap between the
two most luminous galaxies in group halo \citep{Jones2003}, while the
old/young is based on the mass assembly history of the halo \citep{Raouf2014}. It
has been previously shown that a large fraction of old/young galaxy groups are those
with a large/small luminosity gap, signaling some similarities in the evolution
of fossil groups and dynamically old groups.

We find that the BGG major merger phenomenon is an earlier event in relaxed
groups than in unrelaxed groups. Furthermore, the last major merger of BGGs
hosted in young groups occurs at a later epoch than the same process in
fossil and old groups. This is, of course, expected, since galaxy mergers
involve subhalos, that also build up the halos.
Statistically, more than 85\% of young groups have
experienced their last major merger before a look-back time of 1 Gyr, in
comparison to less than 55\% for old groups. The BGG in old and
fossil systems have a lack of recent major mergers in their evolution, in
agreement with earlier observations of fossil groups of galaxies
\citep{Khosroshahi2006b, Smith2010}. This was also shown by
\citet{Von08} who highlight of the same results but with limited
statistics. Our results are also in agreement with the recent study of
\citet{Kundert2017} using the Illustris simulations and the assembly history
of high luminosity gap groups.
 
In the $\Lambda$CDM cosmological framework, one expects that the
group velocity dispersion should be roughly constant since $z=1$.  The evolution of the velocity dispersion
is mainly driven by the evolution of the virial velocity, with radial anisotropy playing a small role. Such an evolution with redshift is affected by mass assembly, halo concentration, and the expansion rate of the Universe.

We obtain interesting results when tracing the group velocity dispersion with
different galaxy populations in different classes of groups.
We find that the velocity dispersion of  old groups traced by very luminous
galaxies decreases in time, except for the most massive groups.
This is a sign of dynamical friction causing the massive galaxies to dissipate
their orbital energy. Since the dynamical friction time scales as the
dynamical time (which depends on position within the virial sphere, but not
on group mass), times a roughly linear function of the ratio of group mass to
galaxy mass, it is too long in the high-mass groups to cause a decrease in
their velocity dispersion traced by these giant galaxies.
Interestingly, we see this effect of dynamical friction in low-mass groups
traced by 
intermediate luminosity galaxies, which, again is caused by the short
dynamical friction time expected for such galaxies.

Another interesting feature is the slight rise of the  velocity dispersion of young
groups of intermediate and high masses. These young groups are acquiring
their mass more rapidly than the average group, and thus their virial
velocity is also rising instead of remaining roughly constant, as is the case
for normal groups.

It is important to note that higher fractions of BGG counterparts (giant galaxies) hosted in old
systems are located in the central regions ($r \sim 0.2 \,\rm Mpc$ from the group
center). This, in turn, shows the effect of an off-set between the BGG and group
center when identifying relaxed and unrelaxed systems.

We use the {\sc Radio-SAGE} semi-analytic model of galaxy formation, perhaps
the only such model offering an observable related to AGN outflows, to compare the relationship between the activity of the central 
supermassive black hole as quantified from the 1.4 GHz radio loudness of the BGG and the
epoch of last major merger in central dominant galaxies hosted by dynamically
old and young galaxy groups. By considering the results of the recent
observational study of \citet{Khosroshahi2017} showing the higher accretion
rate of BGGs in unrelaxed systems, we suggest that  BGGs in groups that formed
recently have higher radio luminosity (for the same stellar mass),
independent of the elapsed time since its last major merger.

Extensions to our current work will focus on the stellar populations of
relaxed and unrelaxed galaxy groups using both observations and cosmological
simulations.

\section*{Acknowledgments} \label{sec:acknow}
We thank the anonymous referee for their constructive comments and suggestions which helped to improve the paper.
The Radio Semi-Analytic Galaxy Evolution (Radio-SAGE) model used in this study is publicly available for download at https://github.com/mojtabaraouf/sage as subgroup of main SAGE repository. The Millennium Simulation databases used in this paper, and the web application providing online access to them, were constructed as part of the German Astrophysical Virtual Observatory (GAVO): http://www.mpa-garching.mpg.de/millennium. We acknowledge the Virgo Consortium for access to this data. 

\appendix
\section{Velocity dispersion model for an isolated halo}
\label{app}

This section describes how to estimate the velocity dispersion of
an isolated halo. We begin by computing the
evolution of the circular velocity at the virial radius, $v_{\rm v}$,
hereafter \emph{virial velocity}.
Given the definition of the virial radius, the virial mass enclosed within it
is
\begin{equation}
  	M_{\rm v}(z) = \frac{\Delta}{2} \frac{H^2(z) r_{\rm v}^3}{G}  \ ,
\label{Mofrvir}
\end{equation}
where $G$ is the the gravitational constant and $\Delta = 200$ here is the
mass overdensity relative to the critical density of the Universe at the virial radius.
Since the circular velocity at the virial radius satisfies
$V_{\rm v}^2=G\,M_{\rm v}/r_{\rm v}$, one obtains with equation~(\ref{Mofrvir}) 
\begin{equation}
	V_{\rm v}^3 = \sqrt{\frac{\Delta}{2}}  G M_{\rm v}(z) H(z)   \ .
	\label{eq:V3v}
\end{equation}

We adopt the characteristic halo mass evolution of
$M_{\rm v}(z) = M_0  \exp[- z/(1+z)]$ \citep{Wechsler2002} and the canonical
evolution of the Hubble constant for a flat Universe with
$H(z) = H_0\,E(z)$, with $E(z) = \sqrt{\Omega_{\rm m}^0(1+z)^3 + 1 - \Omega_{\rm m}^0}$, where
$\Omega_{\rm m}^0$ is the present-day matter density parameter.
Note that the product $E(z)\,M(z)$ varies by less than $\pm2\%$
for $1 > z > 0$, and
hence the virial velocity is roughly constant since $z=1$. This was previously noted
from the evolution of halos obtained with Monte Carlo halo merger trees
\citep{Mamon+12}.

Since halos are well fit by NFW density profile \citep{Navarro+96}, the velocity
dispersion can be simply expressed in terms of $V_{\rm v}$.
According to figure~7 of \cite{Lokas&Mamon01}, the ratio of aperture velocity
dispersion (averaged over a cylinder) to circular velocity at the virial
radius is of order 0.6 to 0.7, depending on halo concentration and velocity anisotropy. For isotropic orbits, this ratio is between 0.62 and 0.66
for reasonable concentrations (\citealp{Mauduit&Mamon07}). For more realistic orbits that are isotropic in the center and become 
gradually mildly radial at large radii, this ratio is between 0.65 and 0.69 for reasonable concentrations \citep{Mamon+13}.
In Figure \ref{fig2:sigma-v}, the expected 1D velocity dispersion is shown as
the black line, with typical 0.25 dex uncertainties in halo mass highlighted as the blue
shaded region assuming that the ratio of aperture velocity dispersion to circular velocity
at the virial radius is 0.65.




\label{lastpage}

\end{document}